\documentclass[twocolumn]{aastex63}
\usepackage{threeparttable}
\usepackage{lineno}

\usepackage{amsmath}
\usepackage{subfigure}
\usepackage{CJK}
\graphicspath{{./}{figures/}}

\begin{document}
\begin{CJK*}{UTF8}{gbsn}
\title{A Population Study on the Radio Emission of Fast Blue Optical Transients}

\author{Jian-Feng Liu (刘建峰)}
\affiliation{Institute of Astrophysics, Central China Normal University, Wuhan 430079, China; \url{yuyw@ccnu.edu.cn}, \url{liuld@ccnu.edu.cn}}
\affiliation{Key Laboratory of Quark and Lepton Physics (Central China Normal University), Ministry of Education, Wuhan 430079, China}

\author[0000-0002-8708-0597]{Liang-Duan Liu (刘良端)}
\affiliation{Institute of Astrophysics, Central China Normal University, Wuhan 430079, China; \url{yuyw@ccnu.edu.cn}, \url{liuld@ccnu.edu.cn}}
\affiliation{Key Laboratory of Quark and Lepton Physics (Central China Normal University), Ministry of Education, Wuhan 430079, China}

\author[0000-0002-1067-1911]{Yun-Wei Yu (俞云伟)}
\affiliation{Institute of Astrophysics, Central China Normal University, Wuhan 430079, China; \url{yuyw@ccnu.edu.cn}, \url{liuld@ccnu.edu.cn}}
\affiliation{Key Laboratory of Quark and Lepton Physics (Central China Normal University), Ministry of Education, Wuhan 430079, China}

\author[0000-0002-9195-4904]{Jin-Ping Zhu (朱锦平)}
\affiliation{School of Physics and Astronomy, Monash University, Clayton, VIC 3800, Australia}

\begin{abstract}

Despite the fact that the progenitor of fast blue optical transients (FBOTs) is still up for debate, FBOTs are sometimes suggested to originate from the core-collapse of ultra-stripped stars and be powered by a spinning-down neutron star. Following this consideration, it is expected that the late-time evolution of the progenitor stars can leave important imprints in the circumstellar material (CSM) of the FBOTs, due to the strong mass loss of the stars. The interaction of the FBOT ejecta with the CSM can drive a long-lasting shock to generate radio emission, which thus enables us to probe the CSM properties through radio observation although such observations are still rare. Within the framework of the magnetar-powered model, \cite{Liu2022} fitted the multi-band optical light curves of 40 FBOTs and, hence, the statistical distributions of the FBOT magnetar and ejecta parameters were obtained. Based on these FBOT population results, we investigate the dependence of the radio emission on the mass-loss rate of the progenitors and evaluate the detectability of radio emission from FBOTs with current and future telescopes.  It is found that the distribution of the peak time and peak luminosity of the emission at 8.4\,GHz are primarily in the regions of $t_{\rm{peak},\nu}=10^{2.12\pm0.63}$\,days and $L_{\rm{peak},\nu}=10^{28.73\pm0.83}$\,erg\,s$^{-1}$\,Hz$^{-1}$, respectively. A joint detection of the Zwicky Transient Facility and Very Large Array could achieve success in about 8.7\% FBOTs of $z\leq1$. Furthermore, if considering a joint observation of the Chinese Space Station Telescope and Square Kilometer Array, this fraction can be increased to about 23.9\%.

\end{abstract}


\section{Introduction} \label{sec:intro}

Fast blue optical transients (FBOTs) are recently discovered and mysterious transient emissions of peak magnitudes of $-16 \lesssim M_{\rm peak} \lesssim -23\,{\rm mag}$ and quite blue colors ($g - r \lesssim -0.2$), whose lightcurves usually rise and decline in a duration shorter than about 10\,days \citep[FBOTs; e.g.,][]{Drout2014,Inserra2019}. Two representative classes of models have been proposed in the literature to explain the FBOT emission, including the central engine models and the shock interaction models, instead of the traditional radioactive decay model for normal supernovae (SNe). Specifically, in the first class, the FBOT emission is produced by a low-mass ejecta mainly powered by a central engine, i.e., a rapidly rotating and highly magnetized neutron star \citep[NS; e.g.,][]{Drout2014,Yu2015,Pursiainen2018} or an accreting black hole \citep[e.g.,][]{Kashiyama2015}. A low-mass ejecta plus a newborn NS system could originate from SN explosions of ultra-stripped progenitor stars \citep[e.g.,][]{Tauris2013,Tauris2015,Tauris2017,Suwa2015,Hotokezaka2017,De2018,Sawada2022} including electron capture SNe \citep[e.g.,][]{Moriya2016,Mor2022}, accretion-induced collapses (AICs) of white dwarfs \citep[WD; e.g.,][]{Kasliwal2010,Brooks2017,Yu2015,Yu2019a,Lyutikov2022}, and even mergers of binary NSs, binary WDs or NS--WD binaries\footnote{WD AICs and binary NS or WD mergers can occur in active galactic nucleus accretion disks and also drive bright magnetar-powered explosions \citep{Zhu2021b,Zhu2021a}.} \citep[e.g.,][]{Yu2013,Yu2015,Yu2019b,Zenati2019}. The common feature of the second class of models is to attribute the FBOT emission to the breakout emission of an outwards propagating shock. The shock could be driven by an SN explosion \citep[e.g.,][]{Chevalier2011,Ginzburg2012,Drout2014,Rest2018,Fox2019,Leung2020,Xiang2021,Pellegrino2022} or by a jet choked in a dense stellar envelope and stellar wind \citep{ Gottlieb2022,Soker2022}.

Besides fitting the observational data of FBOTs specifically, a possible way to examine the different models of FBOTs is to compare them with other explosive phenomena statistically, including superluminous SNe (SLSNe),  broad-lined Type Ic SNe (SNe Ic-BL), and long-duration gamma-ray bursts (lGRBs). First of all, these explosive phenomena all show a preference for exclusively star-forming and low-metallicity host galaxies \citep{Drout2014,Lunnan2014,Kruhler2015,Perley2016,Wiseman2020}, which hints these phenomena could have some connections. Another common feature that needs to be emphasized is that most FBOTs, SLSNe, and fractional lGRB-associated SNe (lGRB-SNe) cannot be explained by the traditional radioactive power model \citep[e.g.,][]{Inserra2019,Liu2022}, which usually requires an ultra-high and even unacceptable-high mass of $^{56}$Ni. On the one hand, SLSNe, in particular, the hydrogen-poor ones are widely believed to harbor a rapidly spinning magnetar, which can power the SN explosion effectively by losing its rotational energy \citep[e.g.,][]{Wheeler2000,Kasen2010,Woosley2010,Piro2011,Inserra2013,Yu2017,Liu2017}. Even for the hydrogen-rich SLSNe that are powered by shock interaction, the kinetic energy of the SN ejecta could still be provided by such a magnetar engine. On the other hand, for lGRBs, although their central engines are usually suggested to be a hyper-accreting black hole \citep{Narayan2001,LiuT2018}, the existence of a long-lived magnetar engine has been widely supported by the observed afterglow flares and plateaus \citep{Dai1998,Zhang2001,Dai2006,Yu2010,Metzger2011,Lv2014}. Thus, it is believe that lGRB-SNe can also be influenced by such a mangetar engine \citep{Yu2017,Zhang2022}.  
Following the above considerations, it is nature to suspect that most of FBOTs, SLSNe, and lGRBs could own a magnetar engine but of different parameters and a united origin model can exist among these phenomena \citep{Yu2017,Liu2022,Hu2023}. In practice, the magnetar engine model can indeed provide good fittings to the observational data of most FBOTs \citep{Yu2015,Hotokezaka2017,Rest2018,Margutti2019,Wang2019,Sawada2022,Liu2022}. 

The different models of FBOTs can be further constrained and discriminated based on the properties of their circumstellar medium (CSM), which is highly related to the mass-loss history of the progenitors of the FBOTs. For example, in the shock interaction models, \cite{Xiang2021} and \cite{Pellegrino2022} found that the mass-loss rate of the progenitors should be up to $\sim 1\,M_{\odot}$\,yr$^{-1}$ in order to account for the rapidly evolving and luminous emission of some FBOTs. However, this result is inconsistent with the limits obtained from the radio observations of FBOTs. The radio emission due to the interaction of the explosive ejecta with the CSM can provide a stringent constraint on the density and density profile of the CSM \citep{Weiler2002}, which is therefore usually concerned in the studies of the explosive phenomena. 
However, so far, successful multi-wavelength observations for FBOTs are still rare. Specifically, only five FBOTs (i.e., AT2018lug, AT2018cow, CSS161010, AT2020xnd, and AT2020mrf) have been detected in radio bands, the last four of which were
also detected in X-rays \citep{RiveraSandoval2018,Margutti2019,Ho2020,Coppejans2020,Bright2022,Yao2022}. The majority of known FBOTs have not been detected in the radio bands. This may in part be due to the fact that the best time for radio observation of most FBOTs had been missed since they were found from archival data. So, it is necessary to estimate when the radio observation should be performed after an FBOT emission and how much sensitivity is required for the telescope.

Very recently, in the framework of the magnetar engine model, \cite{Liu2022} have uncovered the statistical properties of the magnetar engines and the FBOT ejecta, which thus enabled us to further predict the possible distribution of the radio afterglow emission of the FBOTs. In the next section, we briefly describe the conventional model for calculating the radio emission and, with the parameters given by \cite{Liu2022}, present the dependence of the calculated light curves on the uncertain mass-loss rate of the progenitors. In Section \ref{Sec:Population simulation}, we generate a mock population of FBOTs based on a Monte Carlo approach and then evaluate the observational prospects of the FBOT radio emission with future telescopes such as Square Kilometer Array (SKA). Finally,
a summary and discussions are given in Section \ref{Sec:Discussions}.

\section{The model} \label{Sec:Model}
Radio emission following FBOTs is expected to arise from the shock interaction of the FBOT ejecta and its environmental material. The statistical results given by \cite{Liu2022} showed that the ejecta masses ($M_{\rm ej}$) of FBOTs and the initial spin period of the magnetars ($P_{\rm i}$) are anti-correlated with each other and, furthermore, this anti-correlation can naturally extend to the parameter regions of SLSNe, lGRBs, and SNe Ic-BL. Such a universal relation indicates that these different explosive phenomena could have a common origin. The most promising scenario is the core collapse of massive stars, which is supported by the spectroscopic observations of 22 FBOTs published in \cite{Ho2021-ZTF}. Furthermore, as inferred by the $P_{\rm i}-M_{\rm ej}$ relationship, the progenitor stars of these explosions are very likely to interact with a main-sequence star or a compact-object companion \citep{Fuller2022,Hu2023}, which can significantly influence the evolution of the angular momentum and the mass-loss of the progenitor stars. The envelope of the FBOT progenitors is very likely to be stripped drastically before the explosion and thus an intense wind-like environment can be produced.

\subsection{Shock dynamics}
Following the above consideration, on the one hand, as usual adoptions in SN calculations, we assume that the density profile of the FBOT ejecta can have a broken power-law form, with a shallow profile in the inner part and a steep profile in the outer part as \citep{Chevalier1989} 

\begin{equation}
\rho_{\mathrm{ej}}(r, t)= \begin{cases}\zeta_{\rho}\left(\frac{M_{\mathrm{ej}}}{v_{\mathrm{tr}}^{3} t^{3}}\right)\left(\frac{r}{v_{\mathrm{tr}} t}\right)^{-\delta}, & v \leqslant v_{\mathrm{tr}} \\ \zeta_{\rho}\left(\frac{M_{\mathrm{ej}}}{v_{\mathrm{tr}}^{3} t^{3}}\right)\left(\frac{r}{v_{\mathrm{tr}} t}\right)^{-n}, & v \geqslant v_{\mathrm{tr}}\end{cases}
\end{equation}
where $M_{\rm ej}$ is the FBOT ejecta mass and the transition velocity is obtained from the density continuity condition at the interface of the two density profiles, which reads 

\begin{equation}
v_{\mathrm{tr}}=\zeta_{v}\left(\frac{E_{\mathrm{K}}}{M_{\mathrm{ej}}}\right)^{1/2},
\end{equation}
where $E_{\mathrm{K}}$ is the kinetic energy of the ejecta. Under the magnetar mechanism, the kinetic energy of the ejecta would always be primarily derived from the spin-down energy of the magnetar which can be determined by magnetar initial spin period $P_{\rm i}$ \citep{Liu2022}. In our calculations,  the power-law indices are taken as $n=10$ and $\delta=1$, which are typical for core-collapse SNe \citep{Chevalier1989}. The values of the numerical coefficients $\zeta_\rho$ and $\zeta_v$ can be found in Equations (6) and (7) in \cite{Kasen2016}.
On the other hand, the CSM of the FBOTs is considered to have a wind-like density profile as\footnote{A realistic index could somewhat deviate from $-2$, which may lead to a relatively flatter or steeper light curve (e.g., a sharp drop off found from the late-time radio emission of AT2020xnd). It should further be noticed that the possible evolution of the micro-physical parameters can also influence the slope of light curves \citep[e.g., ][]{Bright2022}, which cannot be distinguished from the variation of the density profile. }

\begin{equation}
\rho_{\mathrm{CSM}}(r)=\frac{\dot{M}}{4 \pi v_{\text {wind }} r^{2}},
\end{equation}
where $\dot{M}$ is the mass-loss rate of the progenitor and $v_{\text {wind }}$ is the wind velocity. 

After an FBOT explosion, the rapidly expanding ejecta would plow into the wind-like CSM to drive a long-lasting shock wave. The dynamical evolution of the ejecta-CSM interaction can be described analytically by the self-similar solution developed by \cite{Chevalier1982}. The radius of the shock has a power-law evolution with time as

\begin{equation}
R_{\mathrm{sh}}=R_{\mathrm{tr}}\left(\frac{t}{t_{\mathrm{tr}}}\right)^{\frac{n-3}{n-2}},~{\rm for~}t<t_{\rm tr},
\end{equation}
where $t_{\rm tr}$  is the characteristic timescale of shock dynamical evolution 

\begin{equation}
t_{\mathrm{tr}}=\zeta_{\mathrm{tr}} \frac{M_{\mathrm{ej}}}{\dot{M}} \frac{v_{\mathrm{wind}}}{v_{\mathrm{tr}}} \simeq 1600\,E_{\mathrm{K}, 50}^{-0.5} M_{\mathrm{ej}, -1}^{1.5} \dot{M}_{-4}^{-1} v_{\mathrm{wind}, 8}\,\mathrm{\,days},
\end{equation}
where $E_{\mathrm{K}, 50}=E_{\rm{K}}/10^{50}$\,erg, $M_{\mathrm{ej}, -1}=M_{\rm ej}/0.1\,M_\odot$, $\dot{M}_{-4}=\dot{M}/10^{-4}\,M_{\odot}\,$yr$^{-1}$, $v_{\mathrm{wind}, 8}=v_{\mathrm{wind}} /10^{8}\,$cm\,s$^{-1}$,  and the characteristic radius $R_{\rm {th}}=t_{\rm tr} v_{\rm tr}$. The analytic solution of $R_{\rm sh}$ only holds until $t_{\rm tr}$ when the interacting region reaches down to the inner region ejecta. When $t > t_{\rm tr}$, the density profile of the ejecta follows $\rho_{\mathrm{ej}} \propto r^{-\delta}$, the asymptotic solution is derived by assuming that most of the ejecta is in the dense shell, which yields \citep{Moriya2013}

\begin{equation}
R_{\mathrm{sh}} \propto t^{1/2},~{\rm for~}t>t_{\rm tr}.
\end{equation}

The shock can serve to accelerate electrons to relativistic speeds and amplify the primordial environmental magnetic fields. Consequently, lasting radio emission would be generated through the synchrotron radiation of the relativistic electrons. As a phenomenological description for these micro-physics, we can estimate the energy densities of the electrons and magnetic fields by connecting them with the total energy density of the shocked material by fraction parameters $\varepsilon_{e}$ and $\varepsilon_{B}$ as

\begin{equation}
\begin{gathered}
U_{{e}}=\varepsilon_{{e}} \rho_{\mathrm{sh}} v_{\mathrm{sh}}^{2}, \\
U_{B}=\frac{B^{2}}{8 \pi}=\varepsilon_{B} \rho_{\mathrm{sh}} v_{\mathrm{sh}}^{2},
\end{gathered}
\end{equation}
where $\rho_{\mathrm{sh}}v_{\mathrm{sh}}^2$ represents the total post-shocked energy density, $v_{\mathrm{sh}}$ is the shock velocity, and $\rho_{\mathrm{sh}} \simeq 4 \rho_{\mathrm{CSM}}$ is the mass density of the shocked material  which is given by the Rankine-Hugoniot relations for a strong shock of an adiabatic index $\hat{\gamma}=5/3$.

The shock-accelerated electrons are assumed to distribute initially  as a power law with their kinetic energy as \citep{Bell1978,Blandford1978}

\begin{equation}
\frac{{d} N_{e}}{{~d} \gamma_{e}} \propto (\gamma_{e}-1)^{-p},~\gamma_{e} \geq \gamma_{\min }
\end{equation}
where the power-law index is taken to be $p=3$ which is usually inferred from the radio observation of SNe \citep{Chevalier2006,Maeda2013}.  The minimum Lorentz factor of the shock accelerated electrons reads as

\begin{equation}
\gamma_{\min }=\varepsilon_{{e}} \frac{p-2}{p-1} \frac{m_{{p}}}{2 m_{{e}}}\left(\frac{v_{\mathrm{sh}}^2}{c^2}\right)+1
\end{equation}
where $c$ is the speed of light, $m_{{p}}$ and $m_{{e}}$   are the proton and electron masses, respectively. 
According to the distribution, most of the electrons are non-relativistic the number of relativistic electrons that can emit synchrotron radiation could be estimated by

\begin{equation}
N_{{e}, \mathrm{rel}} \simeq\left(\gamma_{\min }-1\right)^{p-1} \frac{M_{\mathrm{sw}}}{m_{{p}}},
\end{equation}
where $M_{\mathrm{sw}}$ is the mass of shocked material.

\subsection{Synchrotron radio emission}

The radiation loss could play an important role in the process and change the electrons distribution for $\gamma_{{e}}$ above the cooling Lorentz factor $\gamma_{\mathrm{c}}$, which is given by $\gamma_{\mathrm{c}}=6 \pi m_{{e}} c /\left(\sigma_{\mathrm{T}} B^{ 2} t\right)$, where $\sigma_{\mathrm{T}}$ is the Thomson cross-section. The radio spectra produced by synchrotron emission are determined by three characteristic frequencies, including (i) the peak frequency $\nu_{\mathrm{p}} \simeq  2e B /\left(\pi m_{{e}} c\right)$, which corresponds to $\gamma_{\rm e}\sim2$, (ii) the cooling frequency $\nu_{\mathrm{c}}=e B \gamma_{\mathrm{c}}^2 /\left(2 \pi m_{{e}} c\right)$, and (iii) the synchrotron self-absorption (SSA) frequency, which is given by \cite{Yu2019b}, i.e.,

\begin{equation}
\nu_{\mathrm{a}}=\left(\frac{5 e N_{{e}, \mathrm{rel}}}{128 \pi R_{\mathrm{sh}}^2 B} \right)^{2 /(p+4)} \nu_\mathrm{p},
\end{equation}
where $e$ is the electron charge.

For the range of model parameters we are most interested in, the relationship between the magnitudes of the three characteristic frequencies always satisfies $\nu_{\mathrm{p}}<\nu_{\mathrm{a}}<\nu_{\mathrm{c}}$. Therefore, the synchrotron luminosity emitted at a frequency $\nu$ can be formulated as 

\begin{equation} \label{Eq:Lnu_syn}
L_\nu=N_{{e}, \mathrm{rel}}  P_{\max,\nu } \times \begin{cases}\left(\frac{\nu_{\mathrm{a}}}{\nu_{\mathrm{p}}}\right)^{-\frac{p-1}{2}}\left(\frac{\nu}{\nu_{\mathrm{a}}}\right)^{\frac{5}{2}}, & \nu_{\mathrm{p}}<\nu<\nu_{\mathrm{a}} ; \\ \left(\frac{\nu}{\nu_{\mathrm{p}}}\right)^{-\frac{p-1}{2}}, & \nu_{\mathrm{a}}<\nu<\nu_{\mathrm{c}} ; \\ \left(\frac{\nu_c}{\nu_{\mathrm{p}}}\right)^{-\frac{p-1}{2}}\left(\frac{\nu}{\nu_{\mathrm{c}}}\right)^{-\frac{p}{2}}, & \nu>\nu_{\mathrm{c}},\end{cases}
\end{equation}
where $P_{\max,\nu}$ is the peak specific synchrotron emission power of a single electron as

\begin{equation}
P_{\max,\nu }=\frac{m_{e} c^2 \sigma_\mathrm{T}}{3 e} B
\end{equation}
which is independent of the electron Lorentz factor $\gamma_{{e}}$.

In principle, the radio emission presented in Equation (\ref{Eq:Lnu_syn}) would further be suppressed by the free-free absorption (FFA) of the unshocked CSM by a suppression factor as $e^{-\tau_{\mathrm{FFA}, \nu}}$ \citep{Yadlapalli2022}. The optical depth of the FFA is given by

\begin{equation}
\tau_{\mathrm{FFA}, \nu}=\int_{R_{\mathrm{sh}}}^{\infty} \alpha_{\mathrm{FFA}, \nu} \mathrm{d} R
\end{equation}
with \citep{Rybicki1979}

\begin{equation}
\alpha_{\mathrm{FFA}, \nu}=0.018\,T^{-\frac{3}{2}} Z^2 n_i n_e \nu^{-2} g_{\mathrm{ff}} \mathrm{\,cm}^{-1},
\end{equation}
where $T=10^5$\,K is the temperature of the CSM \citep{Chevalier2006}, $Z$ is the charge of thermal ions, $g_{\mathrm{ff}}$ is the free-free gaunt factor, and $n_i$ and $n_e$ are the number densities of ions and electrons, respectively.

\begin{figure}
    \centering
	\includegraphics[width = 1.0\linewidth,trim = 10 10 85 60, clip]{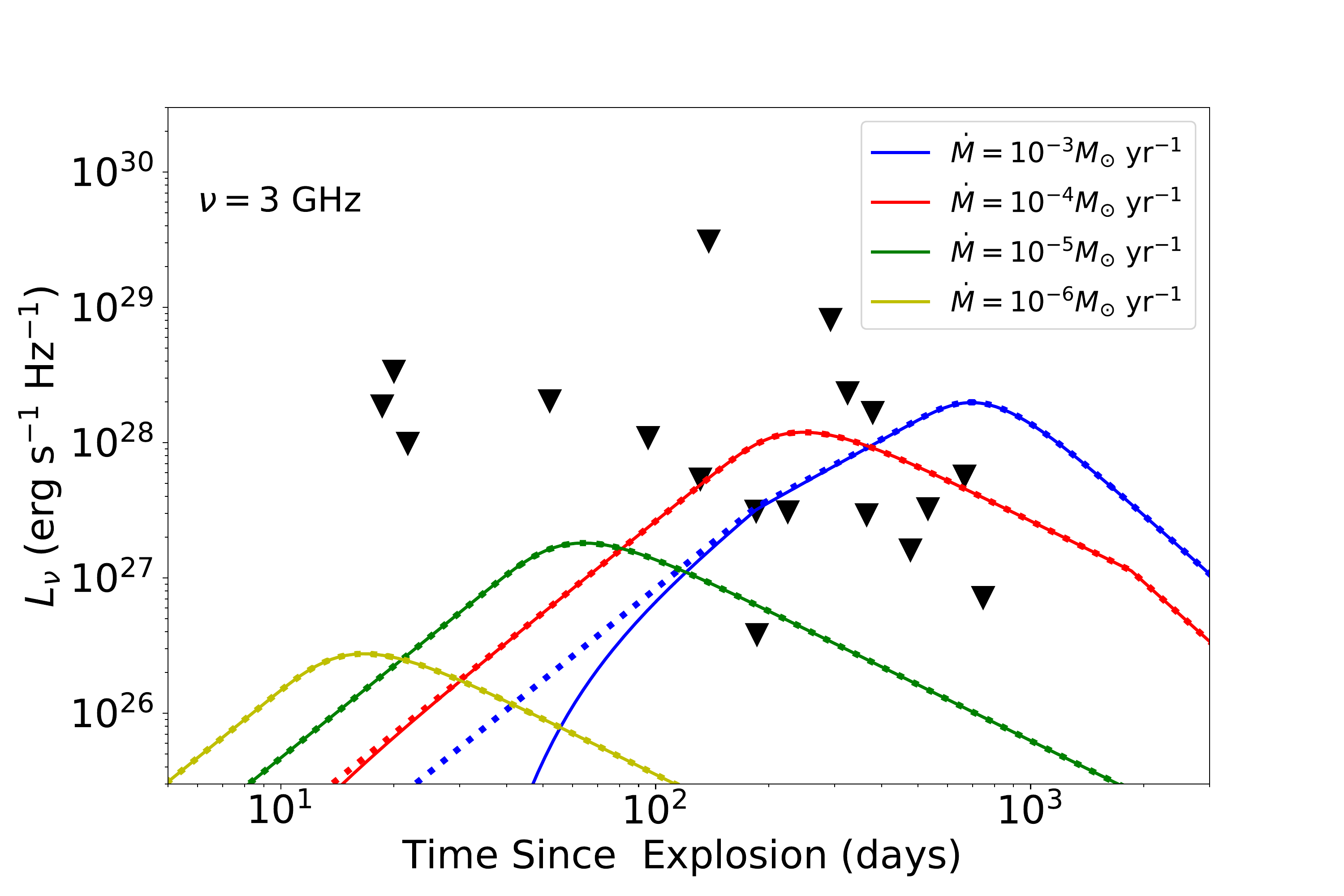}
    \caption{Radio light curves at 3\,GHz with different CSM mass-loss rates $\dot{M}=10^{-6}-10^{-3}\,M_{\odot}$\,yr$^{-1}$, solid and dotted lines indicate that FFA effects are considered and not considered, respectively. 3\,GHz upper limits of FBOTs from VLASS are labeled by downward facing black triangles, the upper limits data are taken from \cite{Ho2021-ZTF}. }\label{fig:Mdots}
\end{figure}

\subsection{Radio light curves} \label{Sec:Results}

The most important implication of the radio emission is the constraint on the mass-loss rate of the FBOT progenitors. Thus, first of all, we investigate the dependence of the radio light curves on the parameter $\dot{M}$, as shown in Figure \ref{fig:Mdots}, where the value of $\dot{M}$  varies from $\dot{M}=10^{-6}-10^{-3}\,M_{\odot}\rm$\,yr$^{-1}$ and the other parameters are taken to be the fiducial values as $M_{\rm ej}=0.11\,M_{\odot}$, $P_{\rm{i}}=9.1{\rm{\,ms}}$, $\varepsilon_B=0.01$, $\varepsilon_{e}=0.1$, and  $v_{\rm wind}=1000$\,km\,s$^{-1}$. It is showed that the FFA can affect the early rising phase of the light curve only for mass-loss rates as high as $\sim10^{-3}\,M_{\odot}\rm$\,yr$^{-1}$. So, we ignore this effect in the following calculations, where only the peak of the light curves are concerned. 

As shown in Figure \ref{fig:Mdots}, larger mass-loss rates result in longer rise times and higher peak luminosity of the radio emission. This result can be understood as follows.
The radio emission peaks when the synchrotron radiation shifts from optically thick to optically thin. This means that radio radiation peaks when the observed frequency $\nu$  and  the SSA  frequency $\nu_{\mathrm{a}}$ are equal. Combined with the dynamic evolution of shock waves, the peak time of radio radiation at frequency $\nu$  is obtained as follows

\begin{equation}
\begin{aligned}
t_{\text {peak},\nu}\simeq &  108 \text{\,days\,} E_{\mathrm{K}, 50}^{0.12} M_{\mathrm{ej}, -1}^{-0.086} \dot{M}_{-4}^{0.58} v_{\mathrm{wind}, 8}^{-0.58} \\
& \times \varepsilon_{B,-2}^{0.34} \varepsilon_{e,-1}^{0.27} \nu_{ 9}^{-0.96},\label{tpeak}
\end{aligned}
\end{equation}
where $\varepsilon_{e,-1}=\varepsilon_{e}/0.1$,  $\varepsilon_{B,-2}=\varepsilon_{B}/0.01$, and $\nu_{9}=\nu/1\,$GHz. 
The corresponding peak luminosity of the emission is

\begin{equation}
\begin{aligned}
L_{\text {peak},\nu}\simeq  & 1.7 \times 10^{27} \operatorname{erg\,s}^{-1} \mathrm{\,Hz}^{-1} E_{\mathrm{K}, 50}^{1.14} M_{\mathrm{ej},-1}^{-0.82} \\
& \times \dot{M}_{-4}^{0.82} v_{\text {wind}, 8}^{-0.82} \varepsilon_{B,-2}^{0.52} \varepsilon_{e,-1}^{0.62} \nu_9^{0.33}.\label{Lpeak}
\end{aligned}
\end{equation}
These expressions show $t_{\text{peak},\nu} \propto \dot{M}^{0.58}$ and $L_{\text{peak},\nu} \propto \dot{M}^{0.82}$, just as found from Figure \ref{fig:Mdots}. Furthermore, Equations (\ref{tpeak}) and  (\ref{Lpeak}) also indicate that the peak time is insensitive to the ejecta mass, whereas the luminosity is highly dependent on the ejecta mass. This means the radio emission of FBOTs would be mainly different from that of typical SNe in the luminosity rather than in the timescale, since the ejecta masses of FBOTs are much smaller than those of typical SNe \citep{Liu2022}.

In Figure \ref{fig:Mdots}, for comparison, we also present the 3 GHz upper limits of 18 FBOTs by the serendipitous observations of the Karl Jansky Very Large Array Sky Survey \citep[VLASS,][]{Lacy2020} from \cite{Ho2021-ZTF} and the upper limits of the flux $F_{\nu}$ at the range of $(109-175)\,\mu$Jy, which correspond to the luminosity upper limits  $L_\nu \sim(0.038-30) \times 10^{28}$\,erg\,s$^{-1}$\,Hz$^{-1}$. Comparing the observations with radio light curves, one  can translate the upper limits to constraints on the density of the CSM around the FBOTs. In some cases where radio upper limits lie above the optically thick locus of the model grid, they can not use to constrain the density of the CSM at all.

\begin{figure*}
    \centering
	\includegraphics[width = 0.8\linewidth,trim = 10 10 50 30, clip]{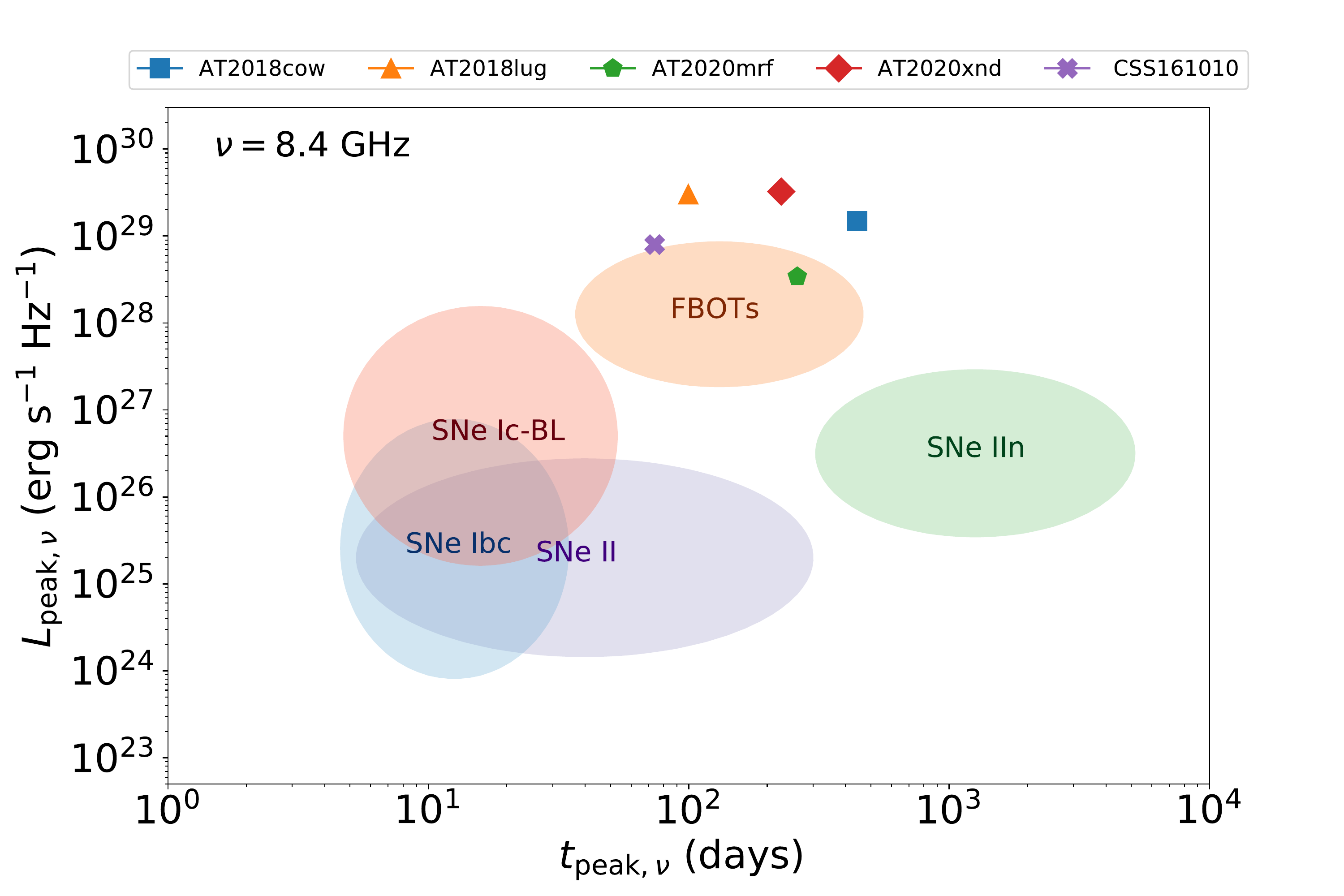}
    \caption{The 1-$\sigma$ region of the peak luminosity and peak time of the 8.4\,GHz radio emission of FBOTs, in comparison with the cases of other types of SNe \cite{Bietenholz2021}. Five detected radio-loud FBOTs are labeled. }\label{fig:LpkTpk}
\end{figure*}

\section{Population simulation} \label{Sec:Population simulation}
\subsection{Method}
Here, for a prospect of the radio observation for FBOTs, we simulate $10,000$ FBOT samples by using a Monte Carlo approach, where each sample is described by a set of parameters $(M_{\rm ej}, P_{\rm{i}}, \varepsilon_e, \varepsilon_B, \dot{M}, v_{\mathrm{wind}}, z)$. Among these parameters, the value ranges of $P_{\rm{i}}$ and $M_{\rm ej}$ have been inferred from the FBOT light curve fittings \citep{Liu2022}, which are $M_{\rm{ej}}=0.11^{+0.22}_{-0.09}\,M_\odot$ and $P_{\rm{i}}=9.1^{+9.3}_{-4.4}\,{\rm{ms}}$. For $M_{\rm{ej}}$ and $P_{\rm{i}}$, we use a two-dimensional Gaussian distribution for sampling. The values of $\varepsilon_B$ and $\varepsilon_e$ can affect the rise time and peak luminosity of the radio emission. We fix $\varepsilon_e=0.1$, which is usually consistent with the radio observations of typical SNe, and assume a log-flat distribution of $\varepsilon_B$ in the interval $[10^{-3},10^{-1}]$. The CSM parameters $\dot{M}$ and $v_{\mathrm{wind}}$ are strongly dependent on the properties of the FBOT's progenitors. For red supergiants, the wind speed has relatively low values as $v_{\rm wind}\sim 10-50$\,km\,s$^{-1}$ and the mass-loss rate is in the range of $\dot{M}\sim 10^{-6}-10^{-4}M_{\odot}$\,yr$^{-1}$. For blue supergiants and Wolf-Rayet stars, the mass-loss rate could be similar with those of red supergiants, but the wind speed can be as high as $v_{\rm wind}\sim 1000-3000$\,km\,s$^{-1}$. The mass-loss rate inferred from the modeling of SNe IIn can even reach an extremely high value of $10^{-3}-10^{-1}\,M_{\odot}$\,yr$^{-1}$. For a comprehensive introduction on the stellar mass-loss, please see the review by \cite{Smith2014}. \cite{Tauris2015} found that the maximum mass-loss rate of ultra-stripped helium stars in binary systems is close to $\sim 10^{-4}M_{\odot}$\,yr$^{-1}$.More recently, \cite{Wu2022} found that mass loss with rates of $\dot{M}\gtrsim10^{-4}\,M_\odot\,{\rm{yr}}^{-1}$ could last for $\sim10^{4}\,{\rm{yrs}}$ until core collapses of close-binary low-mass helium stars, while extreme mass loss $(\dot{M} \gtrsim 10^{-2} M_{\odot}\, \mathrm{yr}^{-1})$ could possibly occur beginning weeks to decades before SN explosions. The CSM interaction of such winds from ultra-stripped stars in binary systems is supposed to be very complicated. More detailed modeling should be
necessary to quantify more accurate late-stage mass transfer rates. Equation (\ref{Lpeak}) shows that the radio luminosity is dependent on $\dot{M}/v_{\rm wind}$ and thus the value of $\dot{M}$ and $v_{\mathrm{wind}}$ cannot be constrained separately by the radio observations. The value of $v_{\rm wind}$ has a large uncertainty, we take the typical escape velocity from a helium star, $v_{\rm wind}=1000$\,km\,s$^{-1}$  \citep{Nugis2000}   and take a log-flat distribution of $\dot{M}/M_{\odot}\rm{\,yr}^{-1}$ in the interval $[10^{-6},10^{-3}]$.

Additionally, in order to compare with the flux sensitivities of the radio telescopes, we need to assign a distance for the simulated samples so that their fluxes can be calculated. Here, we assume the FBOT event rate to trace the cosmic star formation history, which depends on the redshift as follows \citep{Yuksel2008} 

\begin{equation}
f(z)\propto \left[(1+z)^{3.4 \eta}+\left(\frac{1+z}{5000}\right)^{-0.3 \eta}+\left(\frac{1+z}{9}\right)^{-3.5 \eta}\right]^{\frac{1}{\eta}},
\end{equation}
where $\eta=-10$. We sample a uniform distribution of $z$ between 0 and 1.

\subsection{Results}

The properties of the radio light curve at 8.4\,GHz of the different types SNe had been compiled by \cite{Bietenholz2021}, who found that different types of SNe can be characterized by different distributions of $t_{\rm {peak},\nu}$ and $L_{\rm {peak},\nu}$.  According to the 10,000 simulated samples, we find that the distributions of the peak time and luminosity of the 8.4\,GHz radio emission of FBOTs can be roughly described by log-normal functions within ranges of $t_{\rm {peak},\nu}=10^{2.12\pm0.63}$\,days and $L_{\rm {peak},\nu}=10^{28.73 \pm 0.83}$\,erg\,s$^{-1}$\,Hz$^{-1}$. This region is labeled in Figure \ref{fig:LpkTpk} by the orange contours, which is well consistent with the results of the radio-detected FBOTs including AT2018cow. The radio peak luminosities of FBOTs are much higher than those of SNe, while their peak times are briefly comparable. This result can also be understood by Equations (\ref{tpeak}) and (\ref{Lpeak}).

\begin{figure}
    \centering
	\includegraphics[width = 1.0\linewidth,trim = 10 10 70 20, clip]{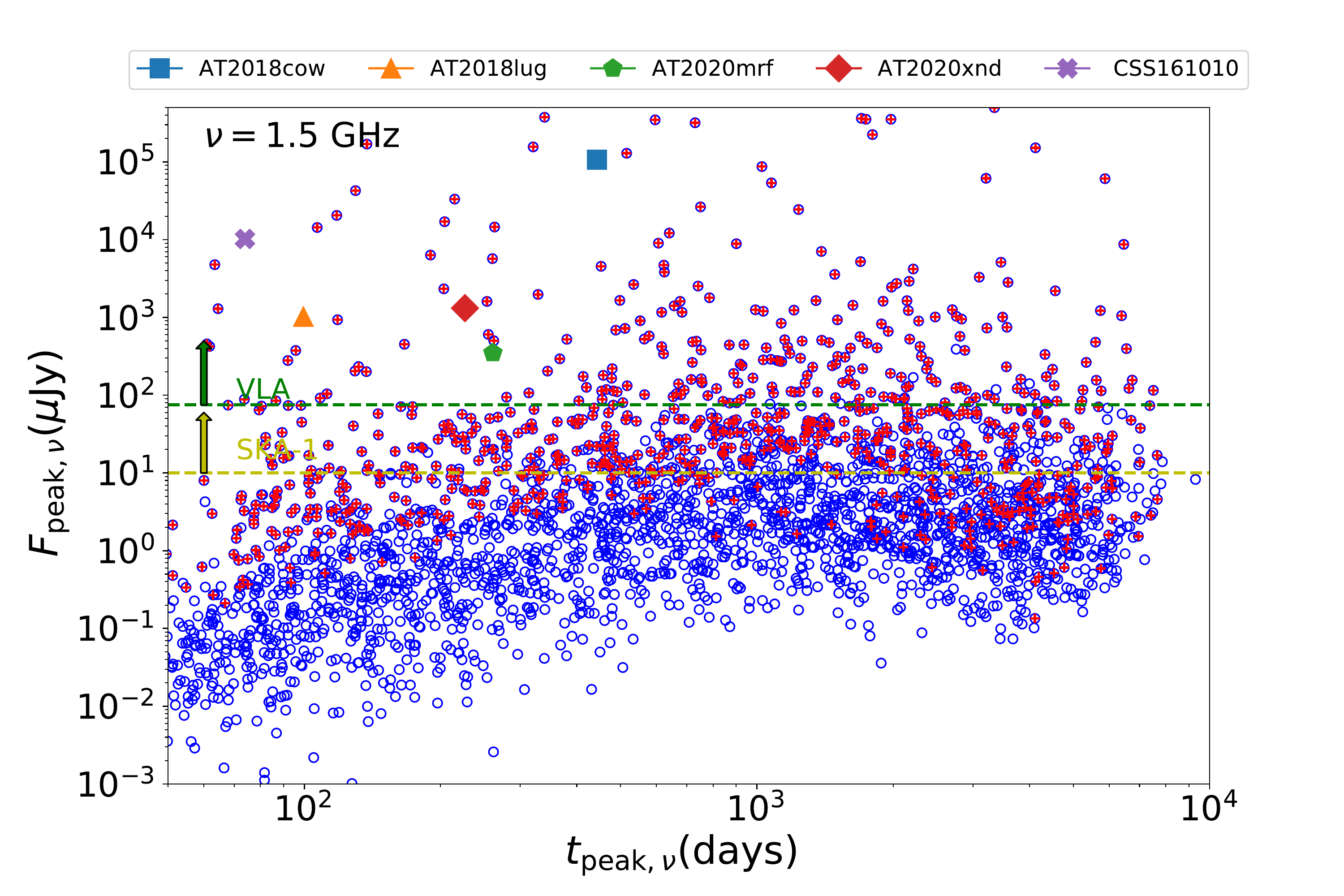}
    \caption{Peak flux $F_{\rm{peak},\nu}$ vs. peak timescale $t_{\rm{peak},\nu}$ of the FBOT radio emission. The blue dots correspond to all simulated samples, and the red dots correspond to those events that are detectable in the optical bands by ZTF.}\label{fig:FpkDLs}
\end{figure}

%\begin{figure}
%    \centering
%	\includegraphics[width = 1.0\linewidth , trim = 10 10 70 20, clip]{Fpk_dist.pdf}	
%    \caption{The distribution of the model-predicted peak flux $F_{\rm{peak},\nu}$.}\label{}
%\end{figure}

To evaluate the detectability of radio emission from FBOTs, we take two typical radio sensitivities to represent the current and future capabilities of radio arrays: VLA with the detection threshold $F_{\rm lim}=75\,\mu$Jy and the phase I of SKA \citep[SKA1,][]{Braun2019} with  $F_{\rm lim}=10\,\mu$Jy at 1.5\,GHz.  In Figure \ref{fig:FpkDLs}, we present the 1.5\,GHz peak fluxes distribution with the peak times for our simulated samples. The blue dots represent our simulated samples, while the red dots indicate the events that can be detected by the Zwicky Transient Facility (ZTF).  

\begin{figure}
    \centering
	\includegraphics[width = 1.0\linewidth,trim = 10 10 80 10,clip]{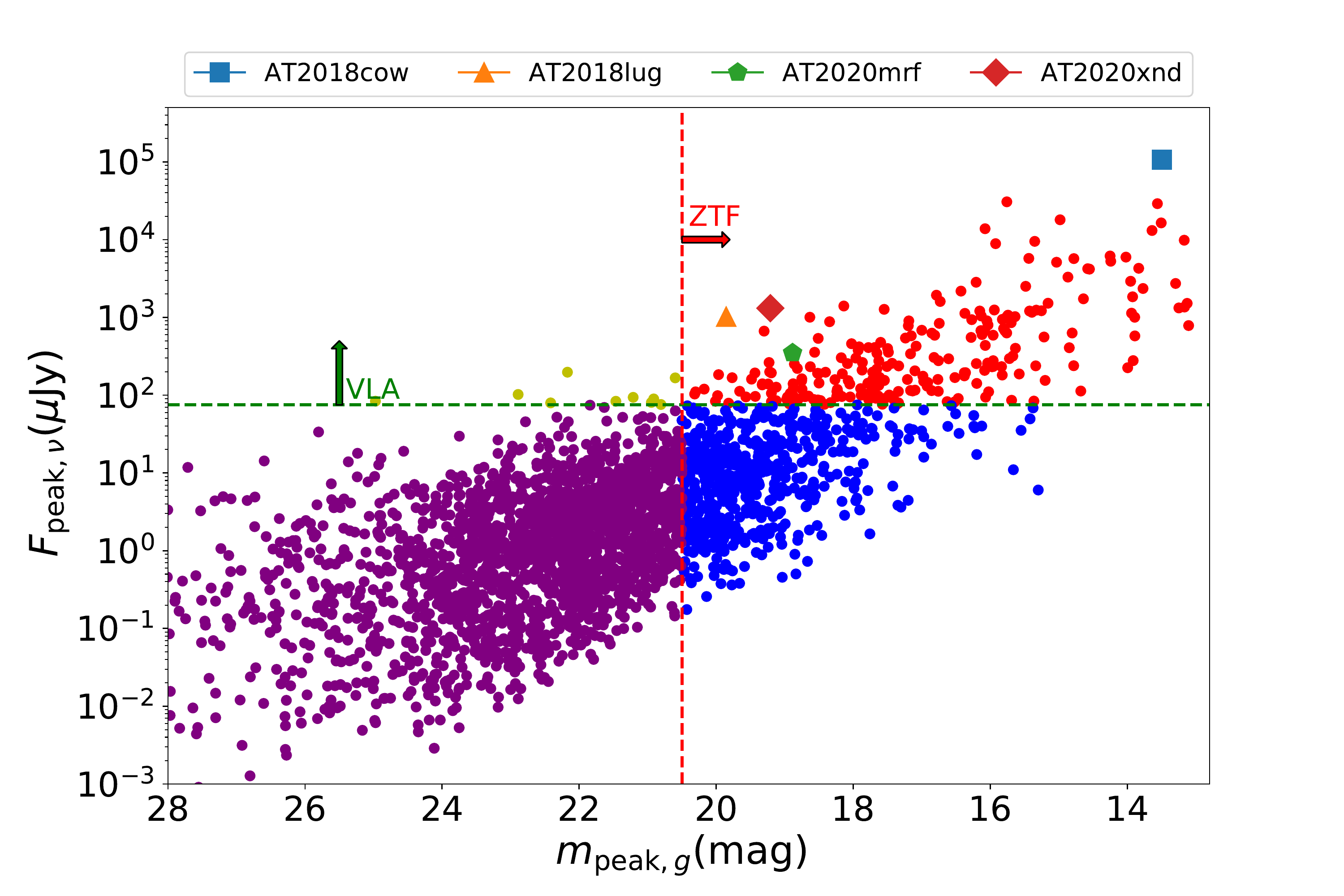}
    \caption{Peak flux in 1.5 GHz $F_{\rm{peak},\nu}$ versus peak apparent magnitude in $g$ band $m_{\rm{peak},g}$. The different detection results of the joint ZTF-VLA observations are labeled by different color dots.}\label{fig:FpkMpk-VLA}
\end{figure}

\begin{figure}
    \centering
	\includegraphics[width = 1.0\linewidth,trim = 10 10 80 10,clip]{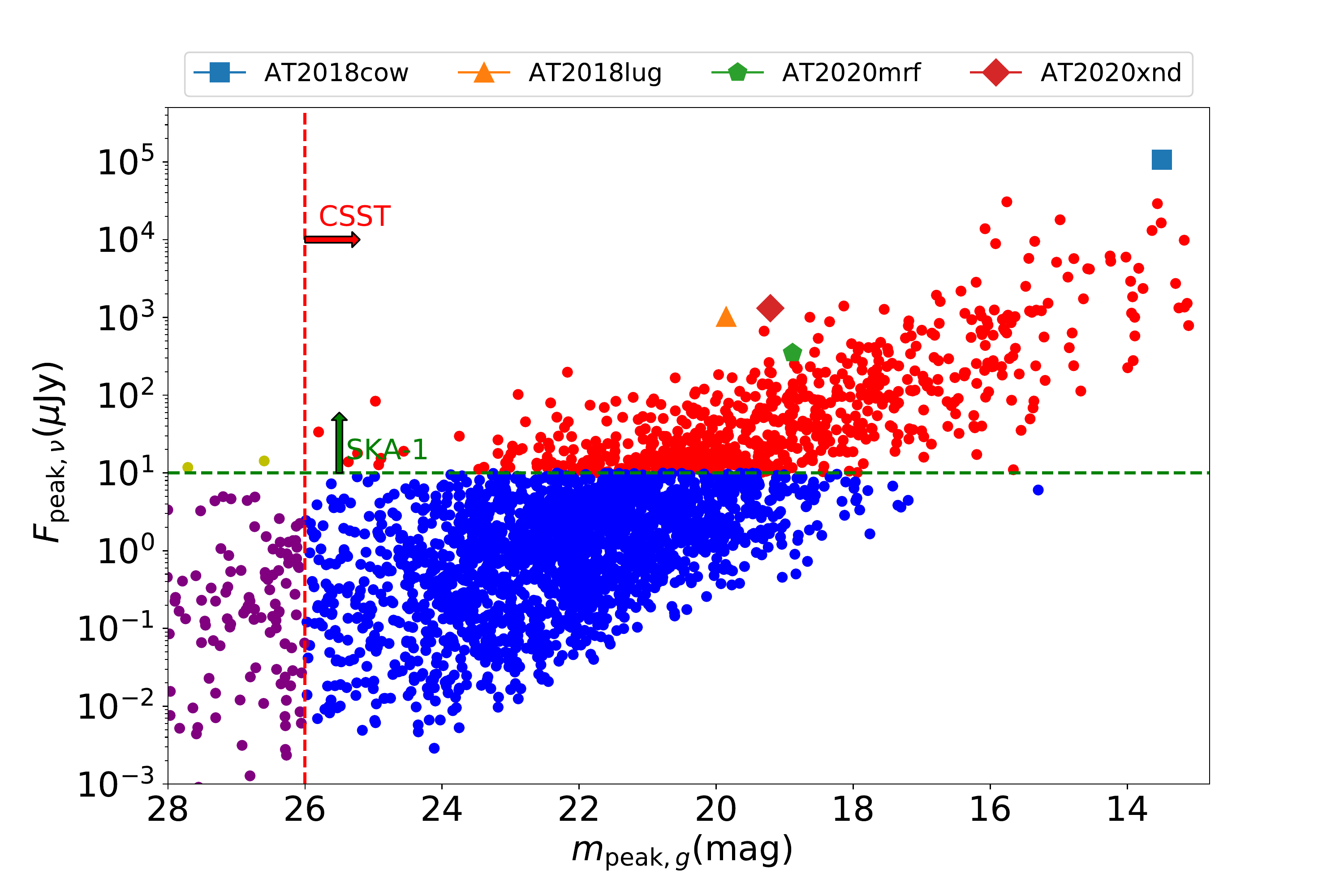}
    \caption{ Same as Figure \ref{fig:FpkMpk-VLA}, but for the joint CSST-SKA1 observation. }\label{fig:FpkMpk-SKA1}
\end{figure}

The distribution of peak flux  $F_{\rm{peak},\nu}$ at 1.5\,GHz and peak apparent magnitude in $g$ band $m_{{\rm peak},g}$  in our simulated samples are presented in Figure \ref{fig:FpkMpk-VLA}.  The different detection results of the joint ZTF-VLA observations are labeled by different color dots.  For the combined ZTF-VLA observations, only 8.72$\%$ of the samples have a chance of being detected simultaneously in optical and radio. Furthermore, for the joint observations of the Chinese Space Station Telescope \citep[CSST,][]{Zhan2011,Cao2018} and SKA1, this fraction can be increased to be about 23.9$\%$, see Figure \ref{fig:FpkMpk-SKA1}.

\section{Summary and discussions} \label{Sec:Discussions}

FBOTs are linked to the explosion of ultra-stripped stars with a very small ejecta mass. The mass-loss histories of the progenitors can be inferred from the constraints on the CSM, which can be obtained by observing the synchrotron radio emission due to the interaction between the ejecta and CSM. Based on these FBOT population results obtained by \cite{Liu2022}, we investigate the prospects of detecting radio emissions from FBOTs. We use a Monte Carlo approach to simulate  $10,000$ radio light curves of FBOTs.  It is found that the distribution of peak time and peak luminosity at 8.4\,GHz satisfy log-normal distributions, with $t_{\rm {peak},\nu}=10^{2.12\pm0.63}$\,days and  $L_{\rm {peak},\nu}=10^{28.73 \pm 0.83}$\,erg\,s$^{-1}$\,Hz$^{-1}$. Compared to radio emission from SNe, the peak luminosities of FBOTs are higher, while the peak timescales are comparable to that of SNe. For a joint observation of ZTF and VLA,  only 8.72$\%$ of the samples have a chance of being detected simultaneously in optical and radio. For the joint observation of CSST and SKA1, the fraction can be increased to 23.9$\%$.

In our calculations, spherical symmetry is assumed for both the ejecta and CSM.  In addition, we assume that the shocked accelerated electrons satisfy a non-thermal power law distribution with their kinetic energy. Recently, \cite{Margalit2021} and \cite{Ho2022} suggested that there is a significant population of electrons remaining in a thermal distribution (i.e., a relativistic Maxwellian) in FBOTs case. Future work may need to more carefully consider the effect of these factors on the FBOT's radio light curves.

This paper focuses on the characteristics of radio emission from FBOTs, while non-thermal X-ray observations of FBOTs can also provide important information. All four FBOTs that were detected in X-ray also detected their radio emissions. The late-time X-ray luminosities of AT2018cow and AT2020mrf are too bright to be an extension of the radio synchrotron spectrum.  Highly variable X-ray emission of FBOTs implicates a central engine, in the form of a compact object \citep{Margutti2019,Yao2022}. Follow-up observations of FBOTs at all wavelengths will likely allow us to obtain more insights into their nature.

Future larger samples or higher sensitivity searches would give meaningful constraints or lead to more detections of radio emission of FBOTs.  This will lead to a better understanding of the progenitors and energy source of FBOTs.

\acknowledgments
We thank the anonymous referee for useful comments, which helped improve this work.
We thank Aming Chen for his insightful comments and suggestions.
This work is supported by the National SKA Program of China (2020SKA0120300), the National Key R\&D Program of China (2021YFA0718500), and the National Natural Science Foundation of China (grant No. 11833003).

\clearpage
\bibliography{FBOT}{}
\bibliographystyle{aasjournal}
\end{CJK*}
\end{document}